\documentclass[12pt,preprint]{aastex}{}
\def\ie{{\it i.e.\/}}
\begin{document}
\title{On the stability of accelerating relativistic
shock waves}
\author{\bf{Giuseppe Palma and Mario Vietri}}
\affil{Scuola Normale Superiore, Pisa, Italy}

\begin{abstract}
We consider the corrugation instability of the self--similar flow
with an accelerating shock in the highly relativistic regime. We
derive the correct dispersion relation for the proper modes in the
self--similar regime, and conclude that this solution is unstable.
\end{abstract}


\section{Introduction}{}

It has long been known that the propagation of a shock through the
outer layers of a star may, for sufficiently steep density
distributions, lead to shock acceleration. Self--similar solutions
(Gandel'man and Frank-Kamenetskii 1956, Sakurai 1960, Raizer 1964,
Grover and Hardy 1966, Hayes 1968) notoriously provide excellent
approximations to the generic solutions, for a reason most clearly
discussed by Raizer (1964): behind the shock, a sonic point is
formed, which prevents causal connection between the region
immediately behind the shock with the more distant regions, which
are sensitive to boundary conditions. Disconnection from initial
conditions limits the number of parameters on which the solution may
depend, while still allowing the existence of a self--similar
solution. Because of the independence from initial conditions, it is
expected that, shortly after their formation, all solutions with
accelerating shocks will quickly approach the self--similar
solution.

Because of the possible relevance of these solutions to the
evolution of Gamma Ray Bursts, there has been recently a strong
surge of interest in their properties, especially in the
relativistic regime, which, at least asymptotically, should be most
appropriate to GRBs. A highly relativistic, self--similar solution
has been presented by Perna and Vietri (2002) in planar geometry,
for which accelerating, self--similar solutions appear for
exponentially decreasing density stratifications outside the star.
In spherical geometry, it is instead possible to find a richer
spectrum of self--similar solutions, for basically any type of
power--law density stratification (Best and Sari 2000, Pan and Sari
2005, Sari 2005). These solutions are often referred to as Type II
solutions (Barenblatt 1996), meaning that the behaviour of the full
solution cannot be determined from dimensional analysis alone, like
in the well--known Sedov solution, but is fixed instead by the
conditions of regularity at some particularly difficult point, like
the sonic point.

The stability of relativistic hydrodynamic solutions with shocks is
rarely considered in the literature. Gruzinov (2001) considered the
linear stability of the spherically symmetric, relativistic
explosions of Blandford and McKee (1976), but nothing is currently
known about the stability of accelerating solutions, in the
relativistic regime. In the Newtonian regime, the self--similar
solutions are known to be unstable, from both a linear analysis
(Chevalier 1990), and a numerical, non--linear one (Luo and
Chevalier 1994). The relativistic solutions differ from this
analysis in two distinct respects: on the one hand, they include
relativistic effects, on the other one, since the shock is supposed
to be highly relativistic, they are obliged to assume the validity
of the ultra--relativistic equation of state, $p = e/3$. Both of
these circumstances lead to a greater susceptibility to
corrugational instabilities. In fact, softening the equation of
state leads to corrugational instability even for shocks propagating
into a {\it constant density} environment: Ryu and Vishniac (1987)
found a range of instabilities provided the adiabatic index is
$\gamma \lesssim 1.2$. Also, assuming an adiabatic index close to
$\gamma =5/3$ leads to both a narrower instability range in the
parameter $k$, and in smaller growth rates than determined by
Chevalier (1990) for $\gamma = 4/3$ in the Newtonian, accelerating
self--similar flow (Vietri, unpublished).

On the other hand, relativistic effects tend to make pressure waves
slow when compared to bulk motion, and thus to decouple nearby layer
of fluid. Thus, if a purely kinematic difference exists in the
motion of adjacent layers, it is more likely to grow because of the
lack of restoring effects.

{}

The plan of this paper is as follows. In the next section, we shall
briefly summarize the properties of the zero--th order solution of
Perna and Vietri (2002). In the following section we shall perform a linear
stability analysis, showing that accelerating, self--similar shocks
are indeed unstable. In the last section, we shall discuss our
results.

\section{The unperturbed solution}

Here we briefly recall the relevant properties of the zero--th order
solution. The self--similar solution for an accelerating
hyperrelativistic shock wave in a planar exponential atmosphere, in
the highly relativistic regime, was derived by Perna and Vietri
(2002). In their notation, the ambient density is given by
\begin{equation}\label{atm}
\rho(x)=\rho_0e^{-k_0x},
\end{equation}
where $\rho_0$ and $k_0$ are constants. If $X^i(s)$ denotes the shock position
at proper time $s$, dimensional and covariance arguments impose
\begin{equation}\label{PV1}
\frac{dX^i}{ds}=\frac{\alpha k^i}{k_{\mu}k^{\mu}s},\qquad i\in\left\{1,2,3\right\},
\end{equation}
where $k_{\mu}\equiv(0,k_0,0,0)$ in the upstream frame and $\alpha$ is a
dimensionless (negative) constant to be determined by imposing a smooth
passage of the flow through a critical point.

Integrating~(\ref{PV1}), it follows that
\begin{equation}\label{PV4}
\frac{k_0t}{-\alpha}=\log\left(\frac{1+V}{1-V}\right)^{1/2}-\frac{1}{V}+\textrm{const},
\end{equation}
whose hyperrelativistic limit is
\begin{equation}\label{PV5}
\frac{k_0(t-t_{\textrm{i}})}{-\alpha}\approx\log\frac{\Gamma}
{\Gamma_{\textrm{i}}}
\end{equation}
($V$ denotes the shock speed, $\Gamma$ its Lorentz factor and the
subscript \emph{i} refers to the initial condition). Moreover,
from~(\ref{atm}),
\begin{equation}\label{PV6}
\frac{\Gamma_{\textrm{f}}}{\Gamma_{\textrm{i}}}\approx\left(
\frac{\rho_{\textrm{f}}}{\rho_{\textrm{i}}}\right)^{1/\alpha}.
\end{equation}

In order to determine the value of $\alpha$, the exact adiabatic
fluid flow equations,
\begin{equation}
\left\{\begin{array}{l}
T^{\mu\nu}_{\quad;\nu}=0\\(n'u^{\mu})_{;\mu}=0
\end{array}\right.
\end{equation}
($T^{\mu\nu}=wu^{\mu}u^{\nu}-pg^{\mu\nu}$ is the energy-momentum
tensor, $w$ and $p$ being respectively local proper enthalpy density and
pressure, $u^{\mu}$ is the fluid four-speed and $n'$ is the baryon
number density as measured in the comoving frame) (Landau and
Lifshitz, 1989), are considered in their highly relativistic limit
given by
\begin{equation}\label{eqenhyp}
\frac{D}{Dt}\frac{e^3}{n'^4}=0;
\end{equation}
\begin{equation}\label{eqmoto}
4\gamma^2e\frac{D
{\mbox{\boldmath{${v}$}}}
}{Dt}+\left(
{\mbox{\boldmath{${\nabla}$}}}
e+
{\mbox{\boldmath{${v}$}}}
\frac{\partial e}{\partial t}\right)=0;
\end{equation}
\begin{equation}\label{PV9}
\frac{\partial n}{\partial t}+
{\mbox{\boldmath{${\nabla}$}}}
\cdot
(n
{\mbox{\boldmath{${v}$}}}
)=0;
\end{equation}
here
{\mbox{\boldmath{${v}$}}}
and $\gamma$ are respectively the fluid speed
and Lorentz factor, and $e=3p=3w/4$ is local proper energy density, where
use of the relativistic equation of state, $p = e/3$ has been made,
as appropriate in the limit of highly relativistic motion. The
operator $D/Dt$ is the usual convective derivative, and $n \equiv
n'\gamma$ is the baryon number density as seen from the upstream
frame, the comoving density being $n'$.

The hyperrelativistic limit of Taub's jump conditions across a
planar shock, with vanishing speed parallel to the shock surface, is
given by
\begin{equation}\label{taub}
n_2=2\Gamma^2n_1,\quad \gamma_2^2=\frac{1}{2}\Gamma^2,\quad e_2=2\Gamma^2w_1,
\end{equation}
where the subscripts 1 and 2 respectively refer to immediately pre-
and post-shock quantities (Landau and Lifshitz, 1989). Upstream
matter is obviously cold in this limit, since its sound speed obeys
$c_s\ll c$; thus, using $p_1\ll e_1\approx\rho_1$, from~(\ref{PV6})
it follows that
\begin{equation}\label{PV12}
e_2=2\frac{\rho_0}{\Gamma_{\textrm{i}}^{\alpha}}
\Gamma^{2+\alpha}\equiv2q_0\Gamma^{2+\alpha}
\end{equation}
and
\begin{equation}\label{PV13}
n_2=2\frac{n_0}{\Gamma_{\textrm{i}}^{\alpha}}
\Gamma^{2+\alpha}\equiv2z_0\Gamma^{2+\alpha},
\end{equation}
where $n_0\equiv\rho_0/m$.

Adopting as self-similarity variable
\begin{equation}\label{PV14}
\xi=k_0(x-X(t))\Gamma^2(t),
\end{equation}
the equations (\ref{taub}), (\ref{PV12}) and~(\ref{PV13}) can be
cast into self--similar form by means of the definitions:
\begin{equation}\label{PV15}
\gamma^2(x,t)=g(\xi)\Gamma^2(t),\quad e(x,t)=q_0R(\xi)
\Gamma^{2+\alpha}(t),\quad n(x,t)=z_0N(\xi)\Gamma^{2+\alpha}(t),
\end{equation}
with
\begin{equation}\label{PV16}
g(0)=\frac{1}{2},\quad R(0)=2,\quad N(0)=2.
\end{equation}

Substituting~(\ref{PV14}) and~(\ref{PV15}) into ~(\ref{eqenhyp}),
(\ref{eqmoto}) and~(\ref{PV9}) it is possible to write equations for
$g$, $R$ and $N$ in the form of a Cauchy problem with ordinary
derivatives (the prime stands for the first derivative with respect
to $\xi$):
\begin{equation}\label{PV17}
R'=\frac{2g\left[-2\alpha(4+\alpha)+(2+\alpha)(\alpha-4\xi)
g\right]R}{\alpha^2+(\alpha-4\xi)g\left[-4\alpha+
(\alpha-4\xi)g\right]}
\end{equation}
\begin{equation}\label{PV18}
g'=\frac{g^2\left[4(\alpha-4\xi)g-14\alpha-3
\alpha^2\right]}{\alpha^2+(\alpha-4\xi)g
\left[-4\alpha+(\alpha-4\xi)g\right]}
\end{equation}
\begin{equation}\label{PV19}{}
N'=\frac{Ng\left\{\alpha^2(18+5\alpha)+2
(\alpha-4\xi)g\left[-2\alpha(5+2\alpha)+
(2+\alpha)(\alpha-4\xi)g\right]\right\}}
{-\alpha^3+(\alpha-4\xi)g\left\{5\alpha^2+
(\alpha-4\xi)g\left[-5\alpha+(\alpha-4\xi)g\right]\right\}}.
\end{equation}

\subsection{Analytical solution}

At the referee's request, we show here how to obtain a (nearly
complete) analytical solution, which may perhaps prove useful.

First of all, we determine $\alpha$, by demanding the simultaneous
vanishing of the numerators and of the denominator of eqs.
\ref{PV17} and \ref{PV18}.  We find:
\begin{equation}
\alpha = -(2+4/\sqrt{3})\approx-4.309401\;.
\end{equation}
Now we introduce a more convenient quantity,
\begin{equation}
Y \equiv (\alpha-4\xi) g(\xi)\;,
\end{equation}
for which we know both the value at the shock front, $Y_0$, and at
the critical point $Y_c$ (from the vanishing of the numerator of eq.
\ref{PV17}, for instance):
\begin{equation}
Y_0 = \frac{\alpha}{2}\;,\;\;\; Y_c = \alpha(2-\sqrt{3})\;.
\end{equation}
From eq. \ref{PV18} we also derive the following differential
equation for $Y$:
\begin{equation}
Y' = \frac{Y}{\alpha-4\xi}
\frac{(2\alpha-3\alpha^2)Y-4\alpha^2}{\alpha^2+ Y^2-4\alpha Y}\;,
\end{equation}
which can easily be integrated to yield the solution in implicit
form:
\begin{equation}{}
\alpha -4\xi = \alpha \exp I_1\;,\;\; I_1 \equiv -4\int_{\alpha/2}^y
\frac{d\!Y}{Y} \frac{\alpha^2+ Y^2-4\alpha
Y}{(2\alpha-3\alpha^2)Y-4\alpha^2}\;,
\end{equation}
where suitable boundary conditions at the shock ($\xi = 0$) have
already been inserted. We also find:
\begin{equation}{}
I_1 = f_1(y)-f_1(\alpha/2)\;,\;\;\; f_1(y) = \frac{4}{\alpha}\left(
\frac{y}{3\alpha-2}+\frac{\alpha}{4}\ln y
-\frac{3\alpha(-4+12\alpha+3\alpha^2)\ln(4\alpha-2y+3\alpha
y)}{4(3\alpha-2)^2}\right)\;.
\end{equation}
If we now compute $I_1$ for $y = Y_c$, we can find an explicit
expression for $\xi_c$, the location of the sonic point. After some
labor, we find:
\begin{equation}
\xi_c = -\frac{1}{2} -\frac{1}{\sqrt{3}} +
\frac{e^{-6+7\sqrt{3}/2}}{\sqrt{3}} \approx -0.462962...\;,
\end{equation}
exactly the same value determined numerically in Perna and Vietri
2002.

Now that we have found the dependence of $Y$ (or of $g$, which
amounts to the same) on $\xi$, we can use eqs. \ref{PV17} and
\ref{PV19} to express $R$ and $N$ as functions of $Y$, using the
highly convenient fact that $R'$ and $N'$ do not depend upon $\xi$
except than through the combination $Y$. We obtain:
\begin{equation}
\frac{dR}{dY} = 2R
\frac{(2+\alpha)Y-2\alpha(4+\alpha)}{(2\alpha-3\alpha^2)Y-4\alpha^2}
\end{equation}
with the obvious solution (with the right boundary conditions at the
shock already inserted):
\begin{equation}{}
R = 2 \exp I_2\;,\;\;\; I_2 \equiv 2\int_{\alpha/2}^Y d\!y
\frac{(2+\alpha)y-2\alpha(4+\alpha)}{(2\alpha-3\alpha^2)y-4\alpha^2}
\end{equation}
from which we easily obtain
\begin{equation}
R = 2\exp\left(2(\frac{7}{2}-2\sqrt{3})(Y-\alpha/2)\right)\;.
\end{equation}
Proceeding analogously for $N$, we find:
\begin{equation}{}
\frac{1}{N}\frac{dN}{dY} =
\frac{\alpha^2+Y(Y-4\alpha)}{(2\alpha-3\alpha^2)Y-4\alpha^2}
\frac{\alpha^2(18+5\alpha)+2 Y\left[-2\alpha(5+2\alpha)+
(2+\alpha)Y\right]} {-\alpha^3+Y\left[5\alpha^2+
Y\left(-5\alpha+Y\right)\right]}\;.
\end{equation}
Including boundary conditions, this has the solution:
\begin{equation}
N = 2 \exp I_3\;,\;\;\; I_3 \equiv \int_{\alpha/2}^Y d\!y
\frac{\alpha^2(18+5\alpha)+-4\alpha(5+2\alpha)y+2(2+\alpha)y^2}
{\alpha(\alpha-y)(-2y+\alpha(4+3y))}\;.
\end{equation}
We also find:
\begin{eqnarray}
I_3 = f_3(Y)-f_3(\alpha/2)\;,\;\;\; f_3(Y) =
\frac{1}{(2-3\alpha)^2\alpha}\left( -2(2+\alpha)(-2+3\alpha)Y
+\right.\nonumber\\
\left.\frac{4\alpha^2(-4-4\alpha+57\alpha^2+9\alpha^3)
\arctan(\frac{3\alpha^2+4Y-6\alpha(1+Y)}{\sqrt{-\alpha^2(2+3\alpha)^2}})}
{\sqrt{-\alpha^2(2+3\alpha)^2}} +\right.\nonumber\\\left.
\alpha(-8+22\alpha+9\alpha^2)\ln(-2Y^2+3\alpha
Y(2+Y)-\alpha^2(4+3Y))\right)\;.
\end{eqnarray}

We remark that this is only a partially analytic solution in any
case, because the relationship between $\xi$ (the radial coordinate)
and $g$ or $Y$ is implicit, making it easy to derive the value
$\xi_c$ of the sonic radius, but making it of little practical
importance anyway.

It is now possible to look at this solution in an instructive
physical way. In fact, one may wonder how a solution may be
accelerating, when a finite amount of energy is available: the
answer is energy concentration. We can compute the parameter
\begin{equation}
{\cal R} \equiv \frac{d\ln E}{d\ln M}
\end{equation}
for, say, all the matter inside the sonic point. It is easy to
check, thanks to the formulae given above, that $E \propto
\Gamma^{2+\alpha}$, while $M\propto \Gamma^\alpha$, both
coefficients of proportionality being of course independent of time.
Thus
\begin{equation}
{\cal R} = 1+\frac{2}{\alpha} \approx 0.535855\;.
\end{equation}
From this we see that the specific energy (per baryon) increases as
$\propto M^{-0.46}$, and thus it increases without bounds when the
total number of baryons between the shock and the sonic point
decreases. At the same time, the total amount of energy in this
layer of matter is dwindling to zero, since it is proportional to
$\propto \Gamma^{2+\alpha} = \Gamma^{-2.309..}$: smaller and smaller
amounts of total energy are used to propel even smaller amounts of
matter.

\section{Perturbation analysis}

There are two ways to carry out the perturbation analysis. Here we
present a simple approach, while in the Appendix we carry out the
full job of perturbing both the equations of hydrodynamics and the
Taub conditions at the shock, putting them together, and finding the
only non-singular solution to the resulting set of equations. The
reason for this double approach is that the first one is commendable
in its simplicity, and is fully satisfactory for the aims of this
paper, while the second one provides some crucial details which are
necessary when comparing the early development of the
numerical solutions to the fully non-linear problem with the linear,
semi-analytic solution; the numerical solutions will be presented
elsewhere, but we can anticipate that these details will come handy
there.

The argument runs as follows. It has been remarked before (Wang,
Loeb and Waxman 2003) that the perturbation problem is not
self-similar, for the following reason. There is an intrinsic
scale-length, $1/k_0$, in the problem, which also determines the
typical {\it transverse} length-scale on which perturbations in the
shocked fluid can travel: this is $1/(\Gamma k_0)$, because the
time-scale for a fluid element moving at speed $1-\Gamma^{-2}/2$ to
cover the distance $1/k_0$ (in the upstream frame) is $1/(k_0
\Gamma)$ in its frame; this implies that the maximum transverse
length coverable by the fluid perturbation is $1/(k_0 \Gamma)$ in
both fluid and observer's frames. The ratio between this transverse
length, and the transverse wavelength $1/k$ of the perturbations is
adimensional, but does depend upon $t$ through $\Gamma$: it thus
breaks the well-known theorem according to which all adimensional
quantities, in self-similar solutions, must be time-independent.

Yet, we can consider two distinct regimes. The first one is that of
short wavelenghts, \ie $\;k/k_0\gg1$. In this case the shock moves
as if in a homogeneous medium, in which case it is well--known
(Anile and Russo 1985) that no instability is present.
Alternatively, we may consider the opposite limit, $ k/(k_0 \Gamma)
\ll 1$, in which case a self-similar perturbation analysis is
warranted: thus, we may consider the case $k=0$ as an approximation
to those late times  when, $\Gamma $ being $\gg 1$, causal phenomena
transverse to the shock's direction of motion cannot carry
disturbances too far. In other words, we neglect the presence of
causal phenomena between crests and valleys, and suppose them to
evolve independently.

Thus, each $y = constant$ slice in the fluid evolves as an
independent solution of the unperturbed equations, with slightly
perturbed constants in the equation for the shock location: the
equation we found above, eq. \ref{PV1}, can be integrated to give,
to most significant orders in $\Gamma$:
\begin{equation}\label{mv}{}
X =t-\frac{\alpha}{k_0}\left(\frac{1}{2\Gamma}\right)^2+
c_1\;,\;\;\;  \Gamma = \Gamma_i \exp(-\frac{k_0t}{\alpha})\;.
\end{equation}
We obtain distinct modes if we perturb the two constants, $\Gamma_i$
and $c_1$, between crests and valleys. When we perturb the quantity
$c_1$ in eq. \ref{mv}, we are leaving the speed difference between
crests and valleys unaffected, and thus $\delta\!X$, the shock
displacement with respect to the unperturbed solution, must be
independent of time. {} Alternatively, we may perturb the quantity
$\Gamma_i$ in eq. \ref{mv}, in which case we find $\delta\!X \propto
\Gamma^{-2}$. These are the two independent modes of the problem.

One can likewise derive the expressions for the perturbed quantities
behind the shock. Let us consider for instance $\delta\! e$ (identical
calculations hold for $\delta\! n$): we use
\begin{equation}
e = 2\Gamma^2\rho(X) R(\xi)\;.
\end{equation}
For the first perturbation, we obviously have $\delta\!\Gamma = 0$,
and thus $\delta\!\xi = -k_0\delta\!X_0\Gamma^2$, from which it
follows that.
\begin{equation}
\delta\!e = 2\Gamma^2\left(-k_0\rho R \delta\!X_0 + \rho R'
(-k_0\delta\!X_0 \Gamma^2)\right) \approx -2k_0\delta\!X_0\Gamma^4 \rho R'\;.
\end{equation}
This equation shows both the time dependence of $\delta\!e$ for this
mode ($\propto \Gamma^4\rho(X) \propto \Gamma^{4+\alpha}$), and its
radial profile $\propto R'$ (while $\delta\!n\propto N'$). We remark that, in this case, the
fractional energy (density) perturbation grows as
\begin{equation}
\frac{\delta\!e}{e}=\frac{\delta\!n}{n} \propto \Gamma^2\;,
\end{equation}
and it thus grows considerably. Analogously,
\begin{equation}
\gamma^2=\Gamma^2g(\xi);
\end{equation}
\begin{equation}
\delta\!\gamma^2=\Gamma^2g'\delta\!\xi=-k_0\delta\!X_0\Gamma^4g'.
\end{equation}

For the other mode, $\delta\!X_0 = 0$, while $\delta\!\Gamma/\Gamma
= \delta\!\Gamma_0/\Gamma_0$. It follows that
\begin{equation}
\delta\!X= \frac{\alpha}{2k_0}\frac{1}{\Gamma^2}
\frac{\delta\!\Gamma_0}{\Gamma_0}\;,
\end{equation}
\begin{equation}
\delta\!\xi = \left(2\xi
-\frac{\alpha}{2}\right)\frac{\delta\!\Gamma_0}{\Gamma_0}\;,
\end{equation}
and, neglecting exponentially small terms, we find:
\begin{equation}
\delta\!e = \left(4\Gamma^2\rho R-\rho R\alpha+2\Gamma^2\rho
R'\left(2\xi-\frac{\alpha}{2}\right)\right)\frac{\delta\!\Gamma_0}
{\Gamma_0} \approx \Gamma^2\rho(4R+4\xi R'-\alpha R')\frac{\delta\!
\Gamma_0}{\Gamma_0}\;.
\end{equation}
From this we again recover both the time dependence of the mode
($\propto \Gamma^2 \rho(X) \propto \Gamma^{2+\alpha}$, which gives
this time $\delta\!e/e$ independent of time), and its space
dependence, $\propto 4R+4\xi R'-\alpha R'$ (identical equations, with
the obvious substitutions, hold again for $\delta\!n$).

For the Lorentz factor, we find:
\begin{equation}
\delta\!\gamma^2=\Gamma^2\frac{\delta\!\Gamma_0}{\Gamma_0}
\left(2g+2\xi g'-\frac{\alpha}{2}g'\right)
\end{equation}


\begin{figure}[!ht]
\begin{center}
\plotone{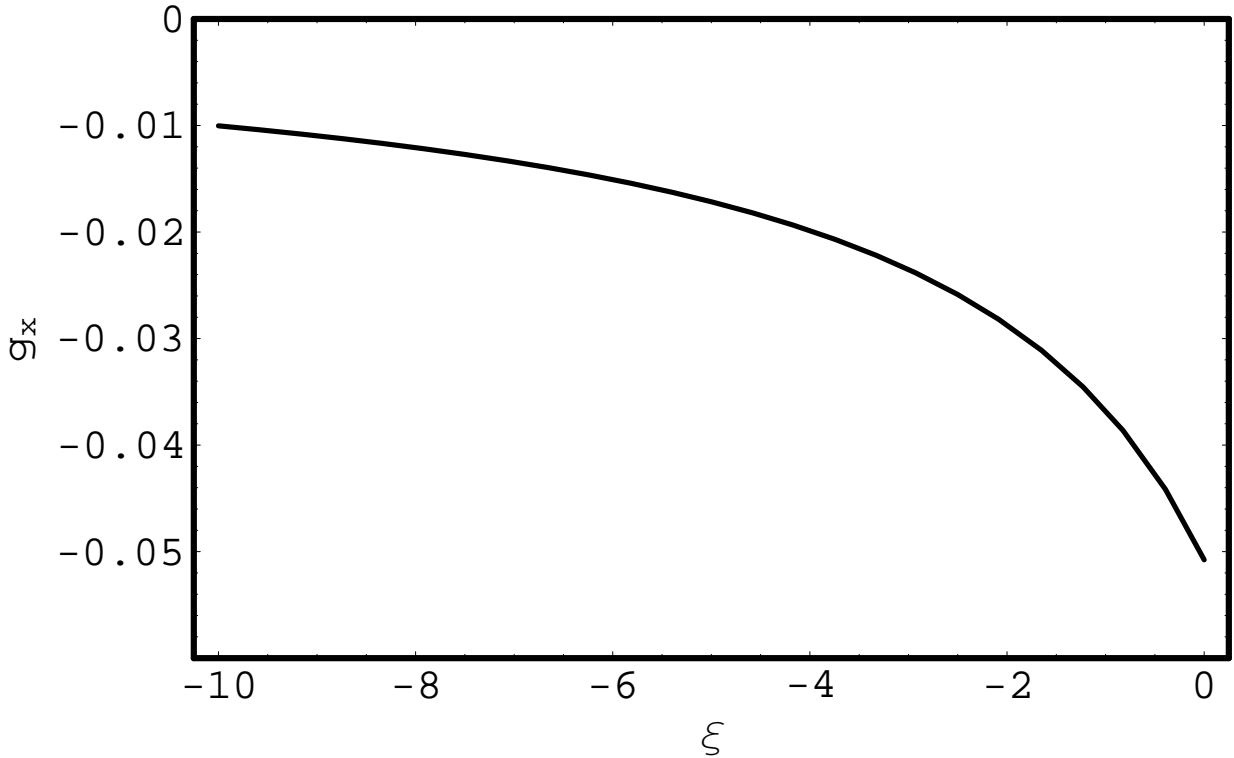} \caption{Spatial dependence of $\delta\!\gamma$,
when $s=3$. In both this figure and in the following ones, we plot
both the result of the analysis given in the text in their
analytical form, and that given by the full analysis given in the
Appendix: that the two superpose within the curve thickness bears
witness to the complete equivalence of the two distinct approaches.}
\label{gx3}
\end{center}
\end{figure}

\begin{figure}[!ht]
\begin{center}
\plotone{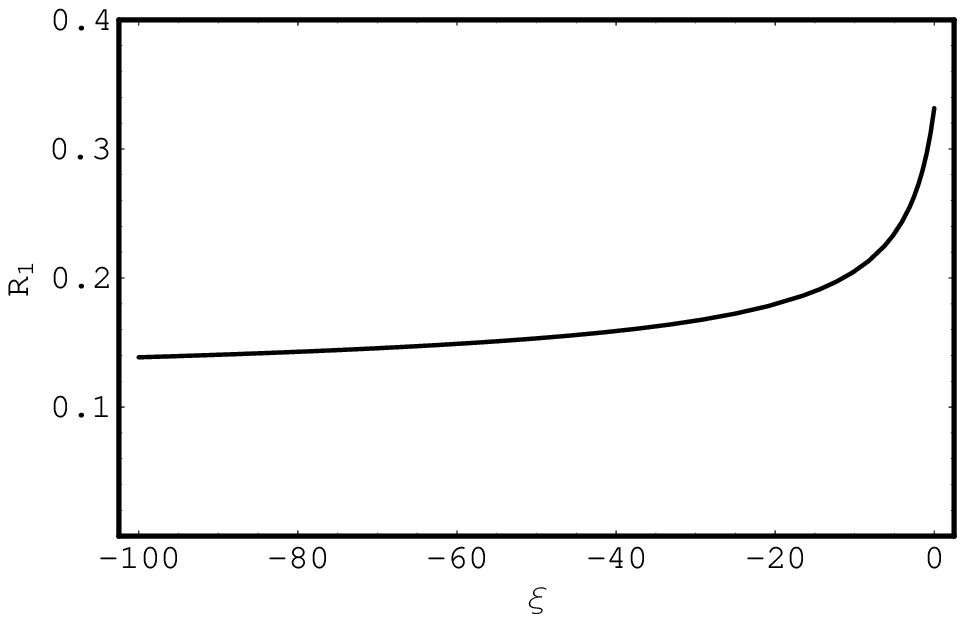} \caption{Spatial dependence of $\delta e$, when
$s=3$.} \label{r13}
\end{center}
\end{figure}

\begin{figure}[!ht]
\begin{center}
\plotone{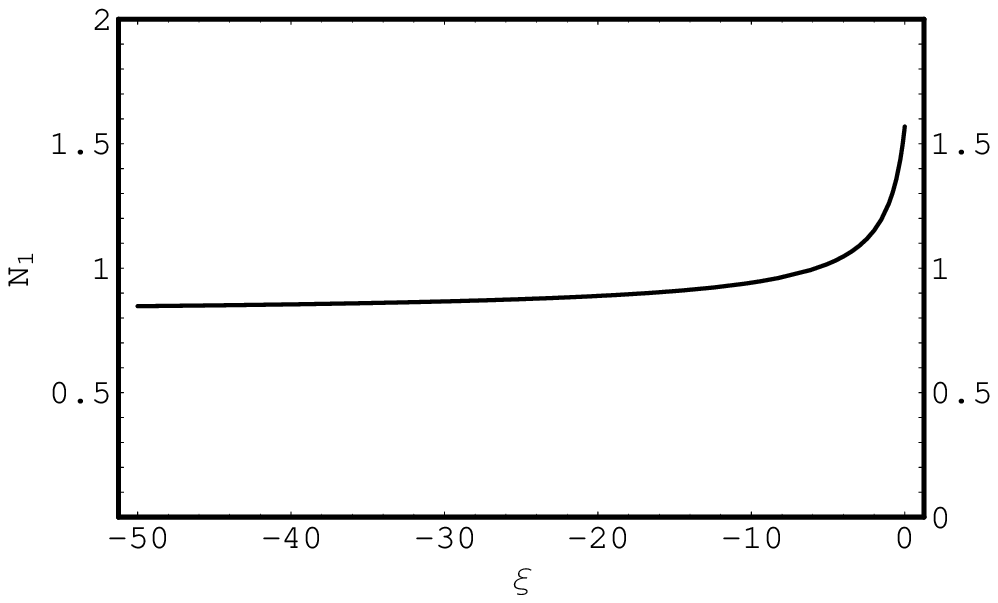} \caption{Spatial dependence of $\delta n$, when
$s=3$.} \label{n13}
\end{center}
\end{figure}

\begin{figure}[!ht]
\begin{center}
\plotone{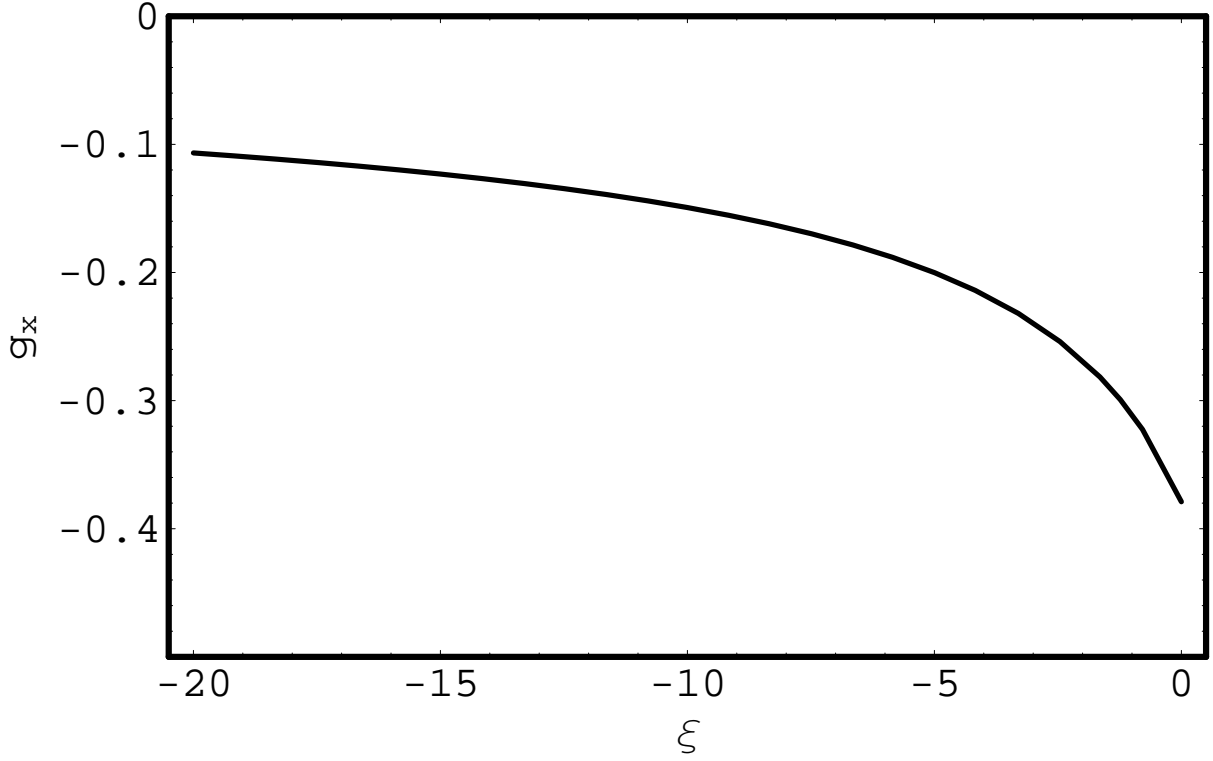} \caption{Spatial dependence of $\delta\!\gamma$,
when $s=1$.} \label{gx1}
\end{center}
\end{figure}

\begin{figure}[!ht]
\begin{center}
\plotone{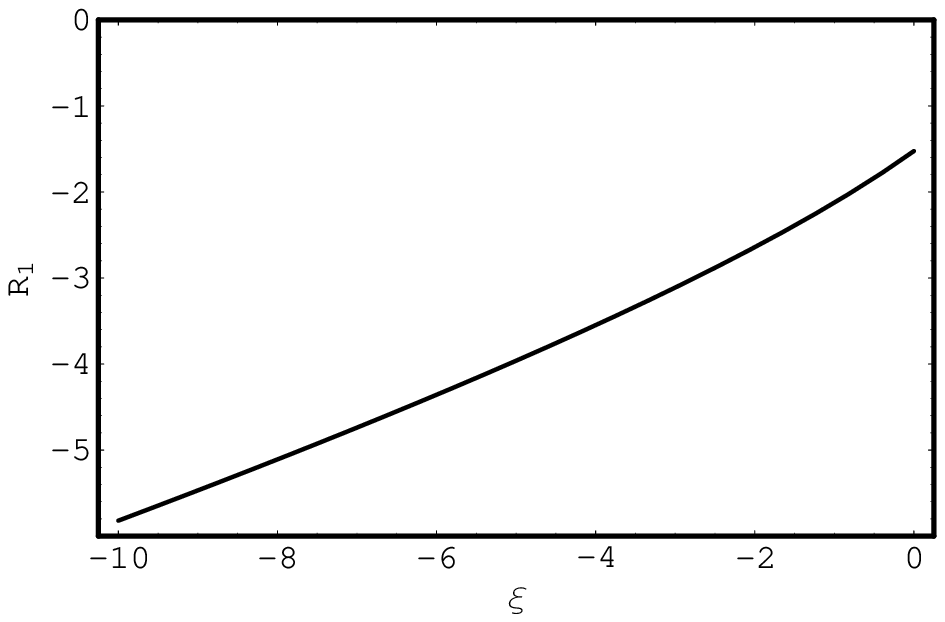} \caption{Spatial dependence of $\delta e$, when
$s=1$.} \label{r11}
\end{center}
\end{figure}

\begin{figure}[!ht]
\begin{center}
\plotone{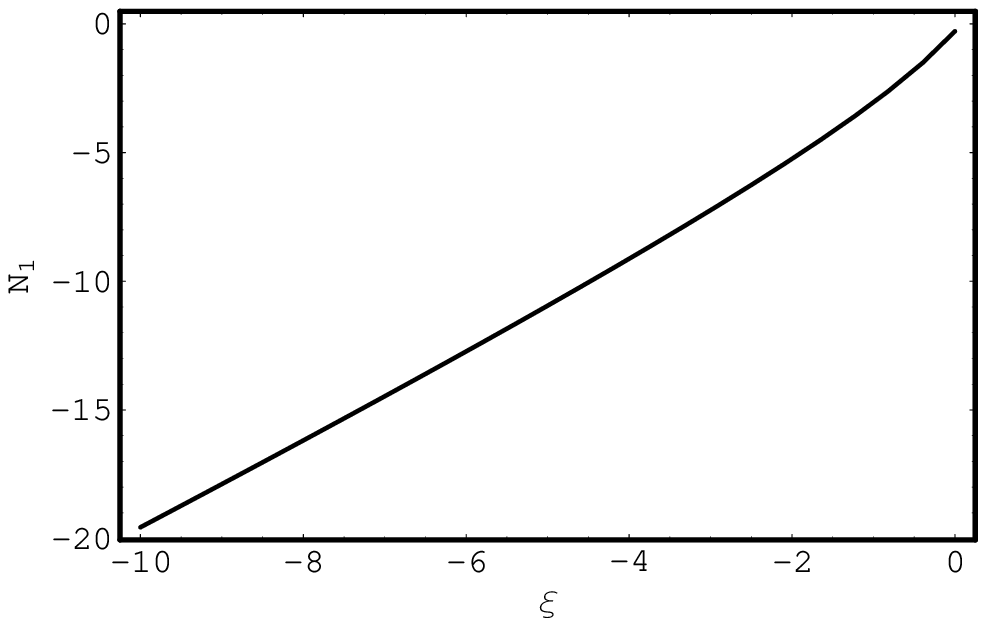} \caption{Spatial dependence of $\delta n$, when
$s=1$.} \label{n11}
\end{center}
\end{figure}

\section{Discussion}

To summarize, we stress here that the hydrodynamical quantities grow
as follows. Let us call $u_x$ the $x$-component of four-velocity.
For $\delta\!\Gamma \neq 0$:
\begin{eqnarray}{}
\delta u_x\approx\delta\!\gamma &\propto& \Gamma^{1} \\
\delta e &\propto& \Gamma^{2+\alpha} \\
\delta n &\propto& \Gamma^{2+\alpha} \\
\delta\!X &\propto& \Gamma^{-2}\;,
\end{eqnarray}
while for the other, stronger mode, $\delta\!X_0 \neq 0$, we find:
\begin{eqnarray}{}
\delta u_x\approx\delta\!\gamma &\propto& \Gamma^{3} \\
\delta e &\propto& \Gamma^{4+\alpha} \\
\delta n &\propto& \Gamma^{4+\alpha} \\
\delta\!X &\propto& \Gamma^0\;.
\end{eqnarray}
Obviously, the most surprising result is that, in this last mode,
the concentrations of energy and baryon number $\delta e$ and
$\delta n$ grow very fast, as $\propto \Gamma^{4+\alpha}$, or,
stated otherwise,
\begin{equation}
\frac{\delta e}{e} \propto \frac{\delta n}{n} \propto \Gamma^2\;.
\end{equation}
Obviously, these concentrations will quickly become nonlinear, and
their subsequent fate can only be ascertained through a numerical
analysis. The same conclusion is reached when one considers the
quick growth of $\delta u_x \propto \Gamma^{3}$.

The linear growth rates are so high that we can expect basically all
small perturbations to become nonlinear during the acceleration of
the ejecta of a Hypernova, when $\Gamma$ changes from
$\Gamma_{\textrm{i}}-1 \ll 1$ to $\Gamma_{\textrm{f}} \approx 100$.

We should remark that the independence of these results from $k$
is a peculiarity of the relativistic solution which does not exist
in the Newtonian counterpart of the problem (Chevalier 1990). This
independence is derived in an explicit way in the Appendix.

Our conclusions differ from those of Wang, Loeb and Waxman (2003),
who were unable to find self--similar perturbations to a similar
problem in spherical geometry, reporting only a marginal growth for
perturbations in the linear regime for an accelerating shock in
spherical symmetry. Though not exactly identical, the growth rates
should however coincide in the limit of large wavenumbers, where
curvature effects become negligible. In support of our work we can
make two different points. On the one hand, there is the the
coincidence of the growth rates when these are computed in two, very
different ways: the simple, physical one (presented above) and the
more detailed, mathematical one to be presented in the Appendix.

On the other hand, there is the similarity between the stability
analysis in the Newtonian (Chevalier 1990) and relativistic regime
(this paper): like us, Chevalier found two independent modes, for
sufficiently small wavenumbers (the limit in which we can compare
directly the two sets of computations), with very simple indices,
$s=0,1$ (but please notice that they are defined in a different way
from ours). Of these, the mode $s=0$ is of course only marginally
unstable, while the other one, corresponding to a time growth
$\propto |t|^{-1}$ for $t\rightarrow 0$, is identified by Chevalier
(1990) as the physically relevant one. This is similar to our result
of two independent modes, with distinct indices $\delta\!e/e \propto
\Gamma^s$, $s=0,2$, the one with $s=2$ resulting in the most severe
(and thus physically relevant) instability.

In a future paper, we will investigate the nonlinear development of
the corrugational instability discussed here.

Thanks are due to the referee, Dr. R. Sari, for helpful comments.

\section*{Appendix}

We present here, in a succinct form, the full perturbation analysis.

A perturbation wrinkling the planar surface of the shock studied
above causes, because of the refraction of the flow lines crossing
an oblique shock, a transverse component (along
{\mbox{\boldmath{$\hat{y}$}}})
 in the shocked matter speed. We perform here
a self--similar analysis of the small perturbations induced by the
shock corrugation. In other words, we take, for the perturbed
quantities:
\begin{equation}\label{P1}
\delta u_x(t,\xi,y)\equiv u_x(t,\xi,y)-u_{x_0}
(t,\xi)=g_x(\xi)\Gamma^s(t)e^{iky}
\end{equation}
\begin{equation}\label{P2}
\delta u_y(t,\xi,y)\equiv u_y(t,\xi,y)=g_y(\xi)
\Gamma^{s+p_1}(t)e^{iky}
\end{equation}
\begin{equation}\label{P3}
\delta e(t,\xi,y)\equiv e(t,\xi,y)-e_0(t,\xi)=q_0
R_1(\xi)\Gamma^{s+\alpha+p_2}(t)e^{iky}
\end{equation}
\begin{equation}\label{P4}
\delta n(t,\xi,y)\equiv n(t,\xi,y)-n_0(t,\xi)=z_0
N_1(\xi)\Gamma^{s+\alpha+p_3}(t)e^{iky}.
\end{equation}
$s$ is the first-order analogous of the 0th-order parameter $\alpha$
and it will be determined by performing a new critical point
analysis. The free parameters $p_1$, $p_2$ and $p_3$ will be
determined shortly. In order to perturb the shock jump conditions,
we will also need a self-similar form for the wrinkle (always in the
upstream frame):
\begin{equation}\label{P5}
\delta\!X(t,y)\equiv X(t,y)-X_0(t)=\frac{\epsilon}{k_0}
\Gamma^{s+p_4}(t)e^{iky}.
\end{equation}
Also $p_4$ will be determined shortly. To simplify the writing of
the equation, we will define $g$ which differs from Perna and
Vietri's:
\begin{equation}
g_{\textrm{new}}\equiv\sqrt{g_{\textrm{old}}}.
\end{equation}

We can also obtain linearized expressions for two useful quantities:
in the hyperrelativistic limit,
\begin{equation}
\delta\gamma(t,\xi,y)=\delta
u_x(t,\xi,y)\left(1-\frac{1}{2\gamma_0^2(t,\xi)}\right)
\end{equation}
\begin{equation}
\delta
{\mbox{\boldmath{${v}$}}} (t,\xi,y)=\frac{1}{\gamma_0(t,\xi)}
\left(\frac{\delta u_x(t,\xi,y)}{\gamma_0^2(t,\xi)}
{\mbox{\boldmath{$\hat{x}$}}} +\delta u_y(t,\xi,y)
{\mbox{\boldmath{$\hat{y}$}}} \right).
\end{equation}

Now we linearize~(\ref{eqenhyp}), (\ref{eqmoto}) and~(\ref{PV9}):
\begin{equation}\label{PV9lin}
\frac{\partial\delta n}{\partial t}+\frac{\partial} {\partial
x}(n_0\delta v_x+v_0\delta n)+\frac{\partial} {\partial y}n_0\delta
v_y=0
\end{equation}
\begin{eqnarray}
&(8\gamma_0e_0\delta\gamma+4\gamma_0^2\delta e) \left(\frac{\partial
v_0}{\partial t}+v_0\frac{\partial v_0} {\partial
x}\right)+\nonumber&\\&4\gamma_0^2e_0\left( \frac{\partial\delta
v_x}{\partial t}+\frac{\partial v_0} {\partial x}\delta
v_x+v_0\frac{\partial\delta v_x} {\partial
x}\right)+\frac{\partial\delta e}{\partial x}
+v_0\frac{\partial\delta e}{\partial t}+\frac{\partial e_0}{\partial
t}\delta v_x=0\label{eqmotoxlin}&
\end{eqnarray}
\begin{equation}\label{eqmotoylin}
4\gamma_0^2e_0\left(\frac{\partial\delta v_y}{\partial t}
+v_0\frac{\partial\delta v_y}{\partial x}\right)+
\frac{\partial\delta e}{\partial y}+\frac{\partial e_0} {\partial
t}\delta v_y=0
\end{equation}
\begin{equation}
\left(\frac{\partial}{\partial t}+v_0\frac{\partial} {\partial
x}\right)\left[-\frac{4e_0\gamma_0^{4/3}} {3n_0^{7/3}}\delta
n+\frac{\gamma_0^{4/3}}{n_0^{4/3}} \delta
e+\frac{4e_0\gamma_0^{1/3}}{3n_0^{4/3}}\delta\gamma\right] +\delta
v_x\frac{\partial}{\partial x}\frac{
\gamma_0^{4/3}e_0}{n_0^{4/3}}=0.\label{eqenhyplin}
\end{equation}
Substitution of~(\ref{P1}), (\ref{P2}), (\ref{P3}) and~(\ref{P4})
into~(\ref{PV9lin}), (\ref{eqmotoxlin}), (\ref{eqmotoylin})
and~(\ref{eqenhyplin}) yields (after some algebra and neglecting
terms of manisfestly inferior order in $\Gamma$):
\begin{eqnarray}
&\Gamma^{p_3-1}g\left\{-2\alpha N_1g'+\alpha gN_1'+
g^3\left[2(p_3+s+\alpha)N_1-(\alpha-4\xi)N_1'\right]
\right\}\nonumber&\\&-2\alpha\left[gg_xN'+N\left
(i\frac{k}{k_0}\Gamma^{p_1}g^3g_y-3g_xg'+gg_x'
\right)\right]=0\label{EQ1gen}&
\end{eqnarray}
\begin{eqnarray}
&4R\left\{g_x\left[2(s-1)g^3+\left((\alpha-4\xi)g^2-3
\alpha\right)g'\right]-g\left[-\alpha+(\alpha-4\xi)
g^2\right]g_x'\right\}\nonumber&\\&+2\alpha gg_xR'+
\Gamma^{p_2-1}g\left\{2R_1\left[(4+p_2+s+\alpha)g^3-
2\left(-\alpha+(\alpha-4\xi)g^2\right)g'\right]
\right.\nonumber&\\&\left.-g\left[\alpha+(\alpha-4\xi)
g^2\right]R_1'\right\}=0\label{EQ2gen}&
\end{eqnarray}
\begin{eqnarray}
&i\frac{k}{k_0}\alpha\Gamma^{p_2-3}g^2R_1-\Gamma^{p_1}
\left\{2R\left[g_y\left(2(p_1+s-1)g^3+\left(-\alpha+
(\alpha-4\xi)g^2\right)g'\right)\right.\right.\nonumber&\\
&\left.\left.-g\left(-\alpha+(\alpha-4\xi)g^2\right)
g_y'\right]+\alpha gg_yR'\right\}=0\label{EQ3gen}&
\end{eqnarray}
\begin{eqnarray}
&2N\left\{4g\left[\alpha+2(\alpha-4\xi)g^2\right]
g_xRN'+N\left[2R\left(g_x\left(2(3s-\alpha-1)g^3
\right.\right.\right.\right.\nonumber&\\&\left.\left.
\left.\left.-\left(5\alpha+(\alpha-4\xi)g^2\right)
g'\right)-3g\left(-\alpha+(\alpha-4\xi)g^2\right)
g_x'\right)\right.\right.\nonumber&\\&\left.\left.
-3g\left(\alpha+2(\alpha-4\xi)g^2\right)g_xR'
\right]\right\}\nonumber&\\&-3\Gamma^{p_2-1}Ng
\left\{2R_1\left[2g\left(\alpha-(\alpha-4\xi)
g^2\right)N'+N\left((4-3p_2-3s+\alpha)g^3\right.
\right.\right.\nonumber&\\&\left.\left.\left.- 2\alpha
g'+2(\alpha-4\xi)g^2g'\right)\right]+
3Ng\left[-\alpha+(\alpha-4\xi)g^2\right]R_1'
\right\}\nonumber&\\&+4\Gamma^{p_3-1}g\left\{
7g\left[\alpha-(\alpha-4\xi)g^2\right]N_1RN'
+N\left[R\left(-2(3p_3+3s-\alpha-4)g^3N_1
\right.\right.\right.\nonumber&\\&\left.\left.
\left.+(4N_1g'+3gN_1')\left(-\alpha+(\alpha-
4\xi)g^2\right)\right)+3g\left(-\alpha+(\alpha-
4\xi)g^2\right)N_1R'\right]\right\}=0.\label{EQ4gen}&
\end{eqnarray}

We now determine $p_1, p_2, p_3$. We know from the non--relativistic
problem (Landau and Lifshitz 1989) that the shock corrugation
introduces three kinds of perturbations into the post--shock flow:
entropy perturbations, vorticity perturbations, and sound waves. The
first arise because, depending on the instantaneous state of motion
of the corrugated surface, matter to be shocked may be faster or
slower than the average flow; this in turn means that, after the
shock, this matter may be hotter or cooler than average, and this
implies entropy perturbations. Also, refraction of flow lines from
oblique shocks produces vorticity perturbations. And lastly,
pressure waves directed away from the shock are all but inevitable.
The amplitude of these perturbations are coupled by the perturbed
shock jump conditions, so that all physical quantities in the
post--shock flow appear as the superposition of three kinds of
perturbations. This simply implies that, in the above equations, we
should choose the three coefficients $p_1, p_2, p_3$ in such a way
that no physical quantity is always negligible. Such situations are
allowed by the above equations, but they describe perturbed flows
where fewer independent perturbations are present. This of course
may occur because the above equations still bear no information
about the shock jump conditions, and thus may potentially refer also
to strictly local perturbations. Inspection of the above equations
shows that the only satisfactory solution where no physical quantity
is completely neglected, as suggested by physical intuition, is:
\begin{equation}
p_1 = -2\;;\;\;\; p_2 = p_3 = 1\;.
\end{equation}
The fact that one solution for three parameters with four equations
to be satisfied can be found bears witness to the soundness of the
idea that the perturbed flow is also self--similar.

Neglecting terms of lower order in $\Gamma$, it is possible to
rewrite the set of differential equations as a 4D-Cauchy problem:

\begin{eqnarray}
&g_x'=\frac{1}{6N^2gR\left[\alpha^2-4\alpha
(\alpha-4\xi)g^2+(\alpha-4\xi)^2g^4\right]}
\nonumber&\\&\times\left\{7g^2\left[-\alpha^2+(\alpha-4\xi)^2
g^4\right]N_1RN'-Ng\left[\alpha+(\alpha-4\xi)g^2
\right]\right.\nonumber&\\&\left[4(\alpha-4\xi)g^2
R(g_xN'+N_1g')-2\alpha R(2g_xN'+5N_1g')-3\alpha
g(R_1N'+N_1R')\right.\nonumber&\\&\left.+g^3
\left(3(\alpha-4\xi)R_1N'+N_1\left(8(1+\alpha)R+
3(\alpha-4\xi)R'\right)\right)\right]\nonumber&\\
&+N^2\left[6(4+\alpha)(\alpha-4\xi)g^6R_1+14
\alpha^2g_xRg'-48\alpha(\alpha-4\xi)g^2g_xRg'
\right.\nonumber&\\&+g^4\left(-3\alpha(7+3s+
\alpha)R_1+10(\alpha-4\xi)^2g_xRg'\right)-
3\alpha^2g(4R_1g'+g_xR')\nonumber&\\&+(\alpha-
4\xi)g^5\left(-6(\alpha-4\xi)R_1g'+g_x\left(
2(6s+\alpha-8)R+3(\alpha-4\xi)R'\right)\right)\nonumber&\\
&\left.\left.+\alpha g^3\left(18(\alpha-4\xi)R_1g'
+g_x\left((20-24s+2\alpha)R+9(\alpha-4\xi)R'
\right)\right)\right]\right\}\label{gx'}&
\end{eqnarray}

\begin{eqnarray}
&g_y'=\frac{2Rg_y\left[2(s-3)g^3-\alpha g'+
(\alpha-4\xi)g^2g'\right]+\alpha g_ygR'- ig^2\frac{k}{k_0}\alpha
R_1}{2gR\left[-\alpha+ (\alpha-4\xi)g^2\right]}\label{gy'}&
\end{eqnarray}

\begin{eqnarray}
&R_1'=\frac{2}{3N^2g^2\left[\alpha^2-
4\alpha(\alpha-4\xi)g^2+(\alpha-4\xi)^2g^4\right]}\nonumber&\\
&\times\left\{-7g^2\left[\alpha-(\alpha-4\xi)g^2\right]^2
N_1RN'-N^2\left[3(3-s+\alpha)(\alpha-4\xi)g^6R_1+
4\alpha^2g_xRg'\right.\right.\nonumber&\\&-
20\alpha(\alpha-4\xi)g^2g_xRg'+6\alpha^2gR_1g
'+g^4\left(6(1+s)\alpha R_1+4(\alpha-4\xi)^2
g_xRg'\right)\nonumber&\\&-2\alpha g^3
\left((4-6s+\alpha)g_xR+3(\alpha-4\xi)R_1g'\right)
\nonumber&\\&\left.+(\alpha-4\xi)g^5g_x\left(2(\alpha-2)R
+3(\alpha-4\xi)R'\right)\right]+Ng\left[-\alpha+
(\alpha-4\xi)g^2\right]\nonumber&\\&\left[4(\alpha-4\xi)
g^2R(g_xN'+N_1g')-2\alpha R(2g_xN'+5N_1g')-3\alpha
g(R_1N'+N_1R')\right.\nonumber&\\&\left.\left.+g^
3\left(3(\alpha-4\xi)R_1N'+N_1\left(8(1+\alpha)R+
3(\alpha-4\xi)R'\right)\right)\right]\right\}\label{R1'}&
\end{eqnarray}

\begin{eqnarray}
&N_1'=\frac{1}{3Ng^2\left[-\alpha^3+5\alpha^2
(\alpha-4\xi)g^2-5\alpha(\alpha-4\xi)^2g^4+(\alpha-4\xi)^3
g^6\right]}\nonumber&\\&\times\left\{7\alpha g^2\left[\alpha^2-
(\alpha-4\xi)^2g^4\right]N_1RN'+\alpha N^2\left[-6(4+\alpha)
(\alpha-4\xi)g^6R_1\right.\right.\nonumber&\\&+4\alpha^2g_xRg'-
24\alpha(\alpha-4\xi)g^2g_xRg'+g^4\left(3\alpha(7+3s+\alpha)
R_1\right.\nonumber&\\&\left.+8(\alpha-4\xi)^2g_xRg'\right)+3
\alpha^2g(4R_1g'+g_xR')\nonumber&\\&\left.+\alpha g^3\left(-18
(\alpha-4\xi)R_1g'+g_x\left((24s-2\alpha-20)R-3(\alpha-4\xi)
R'\right)\right)\right]\nonumber&\\&+Ng\left[2\alpha^2(7+3s+7
\alpha)g^3N_1R+6(1+s+\alpha)(\alpha-4\xi)^2g^7N_1R\right.\nonumber&
\\&-2\alpha(\alpha-4\xi)^2g^4R(g_xN'+N_1g')+6\alpha^2(\alpha-4\xi)
g^2R(4g_xN'+3N_1g')\nonumber&\\&-2\alpha^3R(5g_xN'+8N_1g')-3
\alpha^3g(R_1N'+N_1R')\nonumber&\\&\left.\left.+\alpha(\alpha-4\xi)
g^5\left(3(\alpha-4\xi)R_1N'+N_1\left(-8(2+3s+2\alpha)R+3
(\alpha-4\xi)R'\right)\right)\right]\right\}.\label{N1'}&
\end{eqnarray}
These equations are linear in the perturbations, so their
denominators are completely determined by the 0th-order solution.
This means that it is possible to find roots of these terms without
integrating~(\ref{gx'}), (\ref{gy'}), (\ref{R1'}) and~(\ref{N1'}).
It is straightforward to show that the denominator of~(\ref{gy'})
never vanishes: since $g^2(\xi)\leq1/2$ and $\alpha,\xi\leq0$, it
follows that $-\alpha+(\alpha-4\xi)g^2>0$.

We now remark that both the term common to the denominators
of~(\ref{gx'}) and~(\ref{R1'}) and the denominator of~(\ref{N1'})
vanish at the 0th-order critical point $\xi_{\textrm{c}}$. We have
thus to impose that, by means of a judicious choice of a unique
parameter $s$, the numerators of three equations vanish
simultaneously with their denominators. That this can be done at all
does appear a bit miraculous.

We now turn to the perturbation of the shock jump (Taub) conditions,
which will provide boundary conditions for the numerical integration
of the above perturbed flow equations.

Exactly like in the 0th-order problem, Taub conditions across the
shock provide boundary conditions $g_x(0)$, $g_y(0)$, $R_1(0)$ and
$N_1(0)$. However, perturbations add some complications. \emph{In
primis}, the shock does not have a unique speed
{\mbox{\boldmath{${v_0}$}}} , but its points move along the $x$ axis
with $y$ coordinate dependent velocities:
\begin{equation}\label{vpert}
{\mbox{\boldmath{${v}$}}} (y,t)=
{\mbox{\boldmath{${v_0}$}}} (t)+\dot{\delta\!X}(y,t)
{\mbox{\boldmath{$\hat{x}$}}} .
\end{equation}
Considering $\delta\!X$ as a small perturbation, in the
hyperrelativistic limit\footnote{From now on, this approximation is
always implicitly assumed.}, the perturbed Lorentz factor of the
shock \footnote{The subscript \emph{up} reminds us that it is the
same Lorentz factor with which one, in the shock frame, sees the
upstream fluid incoming.} is
\begin{equation}\label{Gampert}
\Gamma_{\textrm{up}}(y,t)=\Gamma(t)\left[1+\dot{\delta\!X}(y,t)
\left(\Gamma^2(t)-\frac{1}{2}\right)\right].
\end{equation}

Secondly, because of the wrinkle, the shock normal versor
{\mbox{\boldmath{$\hat{n}$}}}
 does not coincide with
{\mbox{\boldmath{$\hat{x}$}}} (such as the tangent versor
{\mbox{\boldmath{$\hat{t}$}}} $\neq$
{\mbox{\boldmath{$\hat{y}$}}} ). It is now possible to determine the
boundary conditions arising from the shock jump conditions. We
consider a frame locally (in the $y_{\textrm{c}}$\footnote{Subscript
\emph{c} evocatively links to the \emph{c}omoving of the frame
characterized by $y_{\textrm{c}}$ with the associated local shock
surface element.} coordinate) comoving with the shock, and denote
quantities in this frame by means of primes; the usual conditions
for the continuity of the fluxes of particle number, momentum and
energy are: ($[f]$ means $f_1-f_2$)
\begin{equation}\label{nn}
\left[{n^n}'\right]\equiv\left[\overline{n}u'_n\right]=0
\end{equation}
\begin{equation}\label{Tnn}
\left[{T'}^{nn}\right]\equiv\left[w{u'_n}^2+p\right]=0
\end{equation}
\begin{equation}\label{Ttn}
\left[{T'}^{tn}\right]\equiv\left[wu'_nu'_t\right]=0
\end{equation}
\begin{equation}\label{T0n}
\left[{T'}^{0n}\right]\equiv\left[w\gamma'u'_n\right]=0;
\end{equation}
here $\overline{n}$ indicates baryonic density as measured in the
frame locally comoving with the fluid; enthalpy density $w$ and
pressure $p$ are always connected to energy density $e$ in the usual
hyperrelativistic way (Landau and Lifshitz 1989). Now we must
connect these quantities with those for which equations~(\ref{gx'}),
(\ref{gy'}), (\ref{R1'}) and~(\ref{N1'}) have been derived.

Remembering the length dilation passing to the comoving frame,
\begin{equation}\label{eta'}
\delta\!X'_{y_{\textrm{c}}}(y,t)=\Gamma_{\textrm{up}}(y_{\textrm{c}},t)\delta\!X(y,t).
\end{equation}
If $\delta\!X\ll k_0^{-1}$,
\begin{equation}\label{n't'}
{\mbox{\boldmath{$\hat{n}$}}} (y_{\textrm{c}})=\left(
\begin{array}{c}
1\\-\left.\frac{\partial\delta\!X'_{y_{\textrm{c}}}(y,t)}{\partial
y}\right|_{y=y_{\textrm{c}}}
\end{array}\right);
\qquad
{\mbox{\boldmath{$\hat{t}$}}} (y_{\textrm{c}})=\left(
\begin{array}{c}
\left.\frac{\partial\delta\!X'_{y_{\textrm{c}}}(y,t)}{\partial
y}\right|_{y=y_{\textrm{c}}}\\1
\end{array}\right).
\end{equation}

Beginning with the calculation of upstream quantities, spatial
components of four-speed $
{\mbox{\boldmath{${u'}$}}} _{1}=-\Gamma_{\textrm{up}}
{\mbox{\boldmath{${v}$}}}$
 are
decomposed into normal and tangential parts:
\begin{equation}\label{un'up}
u'^n_{1}\equiv
{\mbox{\boldmath{${u'}$}}} _{1}\cdot
{\mbox{\boldmath{$\hat{n}$}}} =-\Gamma(1+\dot{\delta\!X}\Gamma^2)
\end{equation}
\begin{equation}\label{ut'up}
u'^t_{1}\equiv
{\mbox{\boldmath{${u'}$}}} _{1}\cdot
{\mbox{\boldmath{$\hat{t}$}}} =-\Gamma\left.
\frac{\partial\delta\!X'_{y_{\textrm{c}}}(y,t)}{\partial
y}\right|_{y=y_{\textrm{c}}}.
\end{equation}
Remembering that atmosphere is stratified, it follows that energy
density is given by:
\begin{equation}\label{e1}
e_1=q_0\Gamma^{\alpha}(1-k_0\delta\!X),
\end{equation}
while proper baryonic density is given by:
\begin{equation}\label{n1}
\overline{n}_1=z_0\Gamma^{\alpha}(1-k_0\delta\!X).
\end{equation}

The Lorentz factor of the downstream fluid, as measured in the
upstream frame, is:
\begin{equation}\label{u0downlab}
u^0_2(t,y_{\textrm{c}})=\gamma(t,0)+\left.
\frac{\partial\gamma(t,x)}{\partial x}\right|_{x=0}\delta\!X
(t,y_{\textrm{c}})+\delta\gamma(t,0,y_{\textrm{c}}).
\end{equation}
In a similar way, indicating with $v_{\textrm{d}_0}(t,x)$ the
unperturbed downstream speed $1-\frac{1}{2\gamma^2(t,x)}$, it is
possible to write the spatial part of the four-speed as:
\begin{equation}\label{uxdownlab}
u^x_2(t,y_{\textrm{c}})=\gamma(t,0)v_{\textrm{d}_0}(t,0)+\left.
\frac{\partial(\gamma(t,x)v_{\textrm{d}_0}(t,x))}{\partial x}
\right|_{x=0}\delta\!X(t,y_{\textrm{c}})+\delta u_x(t,0,y_{\textrm{c}})
\end{equation}
\begin{equation}\label{uydownlab}
u^y_2(t,y_{\textrm{c}})=\delta u_y(t,0,y_{\textrm{c}}).
\end{equation}
Passing to the shock frame (as usual characterized by
$y_{\textrm{c}}$), the four-speed transform as a four-vector:
\begin{equation}\label{u0downsh}
\gamma'_2\equiv u'^0_2=\Gamma_{\textrm{up}}(u^0_2-|
{\mbox{\boldmath{${v}$}}} |u^x_2)
\end{equation}
\begin{equation}\label{uxdownsh}
u'^x_2=\Gamma_{\textrm{up}}(u^x_2-|
{\mbox{\boldmath{${v}$}}} |u^0_2)
\end{equation}
\begin{equation}\label{uydownsh}
u'^y_2=u^y_2.
\end{equation}
Projecting
{\mbox{\boldmath{${u'_2}$}}} on
{\mbox{\boldmath{$\hat{n}$}}} and
{\mbox{\boldmath{$\hat{t}$}}} , it results
\begin{equation}\label{undownsh}
u'^n_2=u'^x_2
\end{equation}
\begin{equation}\label{utdownsh}
u'^t_2=u'^x_2\left.\frac{\partial\delta\!X'_{y_{\textrm{c}}}(y,t)}
{\partial y}\right|_{y=y_{\textrm{c}}}+u'^y_2.
\end{equation}
Analogously, the energy density is
\begin{equation}\label{e2}
e_2=e_0(t,0)+\left.\frac{\partial e_0(t,x)}{\partial x}
\right|_{x=0}\delta\!X(t,y_{\textrm{c}})+\delta e(t,0,y_{\textrm{c}}),
\end{equation}
while the proper baryonic density (it is necessary to divide the
density in the upstream frame by the downstream Lorentz factor
$u^0_2(t,y_{\textrm{c}})$) is given by
\begin{equation}\label{n2}
\overline{n}_2=\frac{n_0(t,0)+\left.\frac{\partial n_0(t,x)}
{\partial x}\right|_{x=0}\delta\!X(t,y_{\textrm{c}})+ \delta
n(t,0,y_{\textrm{c}})}{u^0_2(t,y_{\textrm{c}})}.
\end{equation}
\vspace{1cm}

Now it is possible to substitute the self-similar form for
perturbations into~(\ref{nn}), (\ref{Tnn}), (\ref{Ttn})
and~(\ref{T0n}). Linearizing Taub's conditions and neglecting terms
of manifestly lower order in $\Gamma$, we find (after much algebra)
\begin{equation}\label{tamub}
4\sqrt{2}\left[p_4-s+2(\sqrt{3}-2)\alpha\right]\epsilon-
\left(16g_x(0)-\sqrt{2}N_1(0)\right)\alpha\Gamma^{p_4+3}=0
\end{equation}
\begin{equation}\label{ta1ub}
4(2\sqrt{3}-3)\epsilon+3\left(4\sqrt{2}g_x(0)-R_1(0)\right)
\Gamma^{p_4+3}=0
\end{equation}
\begin{equation}\label{ta2ub}
ik\epsilon+\sqrt{2}k_0g_y(0)\Gamma^{p_4+3}=0
\end{equation}
\begin{equation}\label{ta0ub}
4\left[2(s-p_4)+(11-6\sqrt{3})\alpha\right]\epsilon+
\left(20\sqrt{2}g_x(0)-3R_1(0)\right)\alpha\Gamma^{p_4+3}=0.
\end{equation}
The requirement that~(\ref{tamub}), (\ref{ta1ub}), (\ref{ta2ub})
and~(\ref{ta0ub}) form a non-singular system of equations for the
boundary conditions imposes $p_4=-3$. Solving the system we obtain:
\begin{equation}\label{gx0}
g_x(0)=-\frac{11+6\sqrt{3}-(3+2\sqrt{3})s}{2\sqrt{2}
(7+4\sqrt{3})}\epsilon
\end{equation}
\begin{equation}\label{gy0}
g_y(0)=-\frac{ik\epsilon}{\sqrt{2}k_0}
\end{equation}
\begin{equation}\label{R10}
R_1(0)=\frac{2\left[-27-14\sqrt{3}+(9+6\sqrt{3})s\right]}
{3(7+4\sqrt{3})}\epsilon
\end{equation}
\begin{equation}\label{N10}
N_1(0)=\frac{2\left[-27-16\sqrt{3}+(21+12\sqrt{3})s\right]}
{(3+2\sqrt{3})(7+4\sqrt{3})}\epsilon.
\end{equation}

All the perturbed variables scale with $\epsilon$: we are free to to
set it $=1$ in the following.

The boundary conditions for $g_x$, $R_1$ and $N_1$ do not depend
upon $k$, nor do equations (\ref{gx'}), (\ref{R1'}), (\ref{N1'}).
Thus, the only quantity that does depend upon $k$ is $g_y$, but
since, as remarked above, its denominator never vanishes, it follows
that $s$ cannot be fixed by $g_y$ nor, by implication, by $k$. Thus
$s$ is independent of $k$.

We can thus restrict our discussion to the 3D Cauchy problem for
$g_x$, $R_1$ and $N_1$ and look for a value of $s$ which leads to
\begin{equation}\label{limdes}
\lim_{\xi\rightarrow\xi_{\textrm{c}}^+}N_{g_x}(\xi)=
\lim_{\xi\rightarrow\xi_{\textrm{c}}^+}N_{R_1}(\xi)=
\lim_{\xi\rightarrow\xi_{\textrm{c}}^+}N_{N_1}(\xi)=0.
\end{equation}
Here $N_{g_x}$ means the numerator of the differential equation for
$g_x$, and likewise for $N_{R_1}$ and $N_{N_1}$. Boundary conditions
are given in $\xi=0$, and it is from here that integration process
can start leftwards, until it reaches the critical point
$\xi_{\textrm{c}}<0$ from the right.

\begin{figure}[!ht]
\begin{center}
\plotone{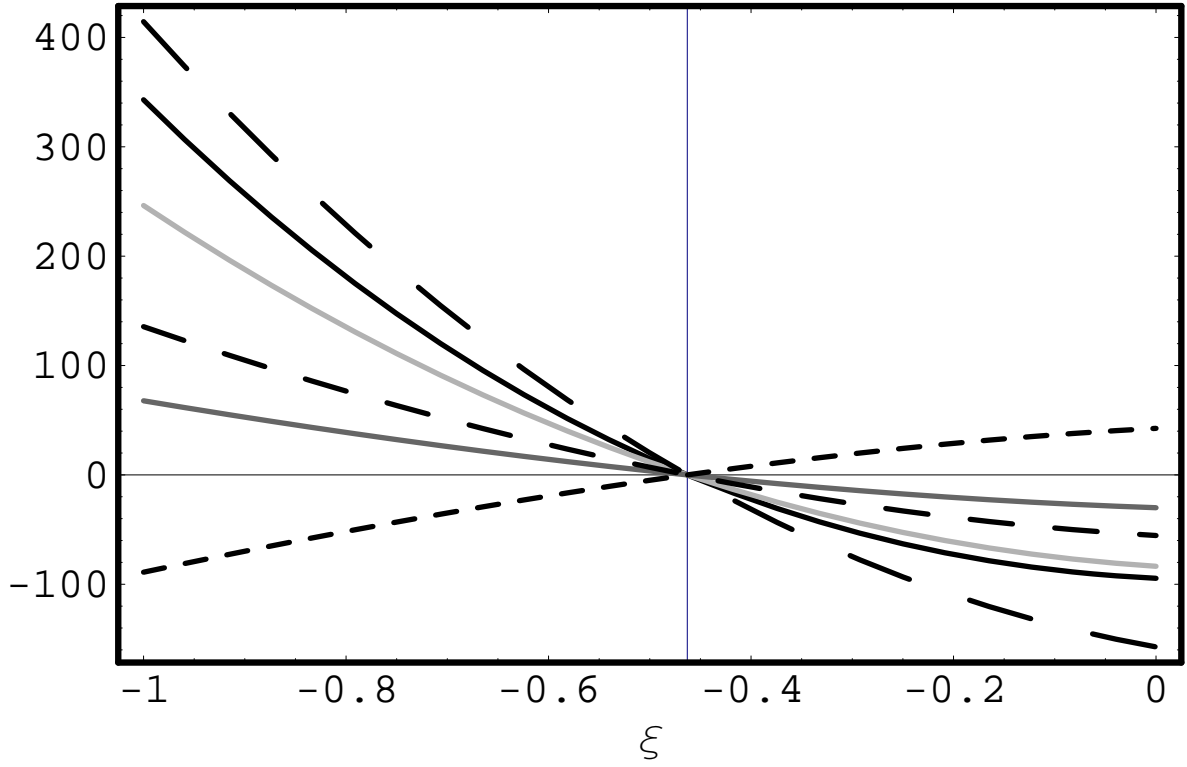} \caption{The run of numerators and denominators of
$g_x'$, $R_1'$ ed $N_1'$ around $\xi_{\textrm{c}}$ for $s=1$.}
\label{crit1}
\end{center}
\end{figure}

\begin{figure}[!ht]
\begin{center}
\plotone{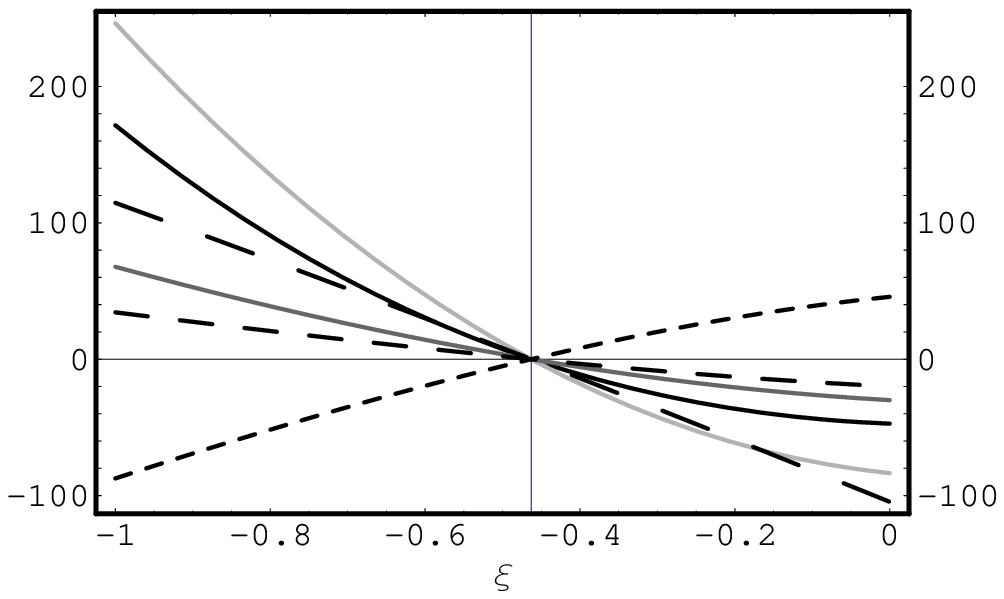} \caption{The run of numerators and denominators of
$g_x'$, $R_1'$ ed $N_1'$ around $\xi_{\textrm{c}}$ for $s=3$.}
\label{crit3}
\end{center}
\end{figure}

Using a binary search based on the direct applications of the
theorem of zeroes, we studied the numerators near $\xi_{\textrm{c}}$
varying $s$; we found these numerators could never vanish for
complex $s$. Thus we investigated these solutions for real values of
$s$. We find two solutions:
\begin{equation}
s_1 = 1\;;\;\;\;\; s_2 = 3\;.
\end{equation}
Integration of the system of equations for $\xi<\xi_{\textrm{c}}$
shows indeed that no divergence is present in our system of
equations: see figures~\ref{crit1} and~\ref{crit3}.

It is also possible to insert the solutions derived in Section 3
into these equations, in order to check that the two sets of
computations are mutually compatible. This has been done by means of
{\it Mathematica}, since the computations are very heavy, and the
expected mutual agreement has indeed been found.


The dependence of all quantities upon the adimensional radius $\xi$
is shown in the figures, first for $s=3$, and then for $s=1$ ($g_y$
figures are obviously reported here as a result of this only section). It
can easily be checked that we have indeed found two self--similar,
distinct solutions, passing without divergence through the sonic
point.

\begin{figure}[!ht]
\begin{center}
\plotone{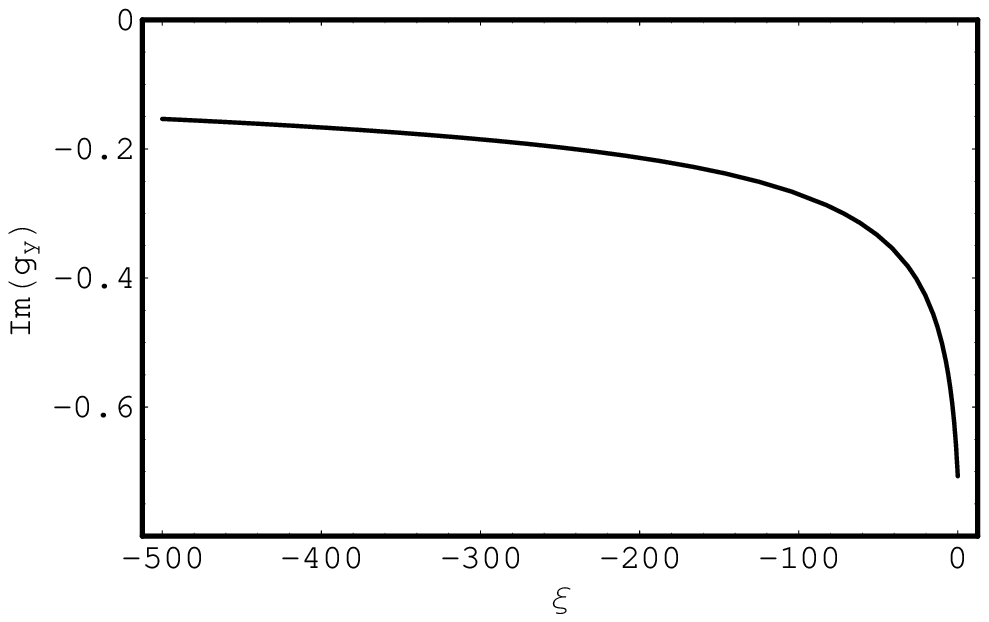} \caption{The imaginary part of $g_y$, when $s=3$ and
$k=k_0$.} \label{gy3}
\end{center}
\end{figure}

\begin{figure}[!ht]
\begin{center}
\plotone{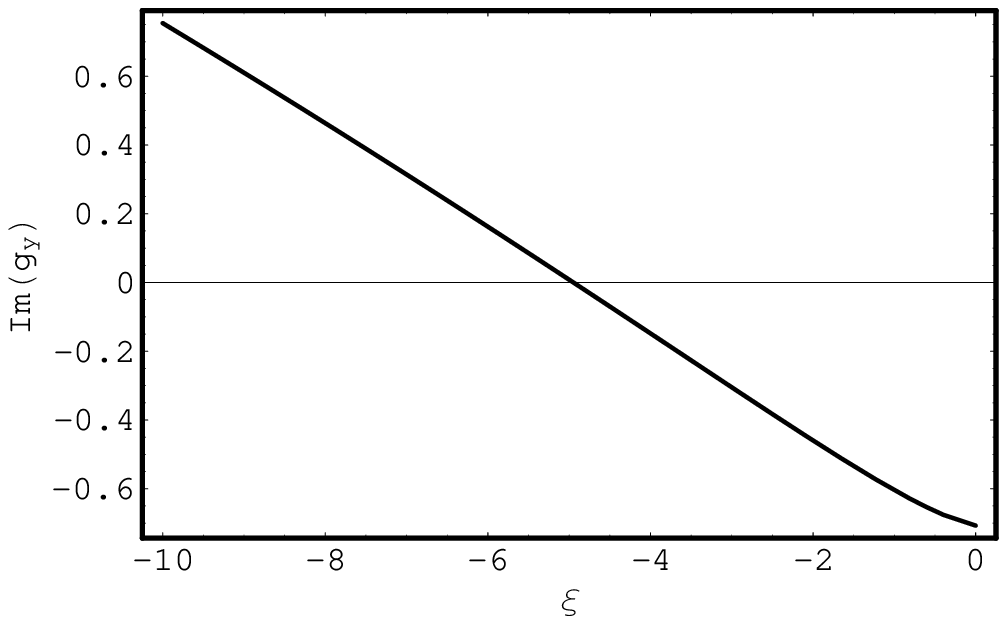} \caption{The imaginary part of $g_y$, when $s=1$
and $k=k_0$.} \label{gy1a}
\end{center}
\end{figure}

Since $s$ is real, so are $g_x$, $R_1$ and $N_1$. However, $g_y$,
the only function depending on $k/k_0$, is purely imaginary: in
fact, the boundary condition is purely imaginary and its
differential equation contains only real terms multiplied by $g_y$
and a term containing $R_1$ multiplied by $i$. Furthermore, $\imath
k/k_0$ enters $g_y(\xi)$ simply as a multiplicative factor. In a
more intuitive way, let us consider several wrinkles of the same
amplitude $\epsilon/k_0$, but with different wave-lengths. Clearly,
the tangential component of the speed at the shock (which, after
all, is the quantity that determines $g_y$) is proportional to
$\sin\arctan\epsilon k/k_0$; as a wrinkle is a small perturbation,
this factor reduces to $\epsilon k/k_0$, thus justifying previous
mathematical result.

\end{document}